\documentclass[12pt]{article}
\usepackage{latexsym}
\usepackage{epsfig}
\usepackage[]{graphicx}
\usepackage[english]{babel}
\usepackage{amsmath}
\usepackage{hhline}
\usepackage{amssymb}
\usepackage{times}
\newcommand \pomeron {I\!\!P}
\begin{document}
 \title {On the universality of cross sections of
hadron -hadron(nuclei) collisions at superhigh energies}
\author {L.~ Frankfurt\\
\it School of Physics and Astronomy,\\
\it  Tel Aviv University,
Tel Aviv, 69978 , Israel\\
M. ~Strikman\\
\it Department of Physics,
Pennsylvania State University,\\
\it University Park, PA  16802, USA\\
M.~Zhalov\\
\it St. Petersburg Nuclear Physics Institute, Gatchina, 188300
Russia}
\date{}
 \maketitle

\begin{abstract}
We analyze the pattern of the onset of complete absorption (the
black limit) in the high energy hadron-hadron collisions. The
black limit arises due to the hard and soft interaction dynamics
as a function of the impact parameters b. Both hard and soft
mechanisms provide universal dependence of the partial amplitude
of the high energy elastic hadron- hadron scattering on the impact
parameter ${\bf b}$ and result in the radius of interaction
 proportional to $\ln (s/s_0)$. We find that with increase of
the collision energies hard interactions lead to a faster increase
of the impact parameter range where the partial wave amplitudes are
approaching the unitarity limit. Consequently,
 we argue that at super high energies when
the radius of hadronic interactions
significantly exceeds static radii of
the interacting hadrons(nuclei) the ratio of total
cross sections of nucleon-nucleon, meson-nucleon, hadron-nucleus,
nucleus-nucleus
collisions  becomes equal to
 one. The same universality is also expected for the
structure functions of nuclei:  $F_{2 A}(x,Q^2)/F_{2 N}(x,Q^2)\to
1$, at  very small $x$, and for the ratio $\sigma_{\gamma A}/\sigma_{\gamma p}$
at superhigh energies. We analyze how
accounting for the energy dependence of the interaction radii
changes the geometry of hadron-nucleus and nucleus-nucleus
collisions, the energy dependence of total, absorption and
inelastic cross sections, the distribution over the number of
wounded nucleons in proton-nucleus collisions and find that
these effects are noticeable  already for the LHC energies and even more so
close to the  Greisen-Zatsepin-Kuzmin  limit.
\end{abstract}

\section{Introduction}

High energy behavior of the hadron-hadron interaction is a subject
of the extensive theoretical studies \cite{heamp} for almost fifty
years. Still many long standing questions such as the universality
of the hadronic and hadron-nucleus total cross sections \cite{unics}
and the relative importance of hard and strong interactions as a function
of the impact parameter remain open.

V.Gribov \cite{Gribovslope} demonstrated long ago that the $t$
channel unitarity of the S-matrix combined with analytic
properties of the scattering amplitude in momentum transfer space
 is
inconsistent with the radius of a hadron being independent of the
incident energy. An assumption that the single pole singularity in
the angular momentum plane (a Pomeron pole) determines high-energy
behavior of the elastic amplitude \cite{Gribovradius,Chew}
predicts an increase of the interaction radius with increase of
energy. This phenomenon leads to increase of the slope parameter
$B_{pp}$ in the t-dependence of the $pp$ elastic cross section
which is by now well established experimentally
 -see \cite{shrinkage} and references therein. Estimates
within the Pomeron pole model show that the radius of interaction
becomes comparable with the mean internucleon distances in nuclei
already at the
LHC energies and even exceeds these distances at energies
corresponding to the Greisen-Zatsepin-Kuzmin  (GZK)
limit \cite{gzk}. This leads to a gradual change of geometry in
hadron-hadron, hadron-nucleus and nucleus-nucleus collisions.
Also, it results in a gross difference between structure
 of the final states in central collisions (for example triggered
by $x_T\equiv 2p_{\perp}/\sqrt{s} \geq 0.01$ dijet production)
and peripheral collisions
at the Tevatron energies and
above \cite{FSW}.

Another fundamental property of the strong interaction is a rather
rapid  increase of the total hadron-hadron cross section  with
energy. Such a behavior is well described, for example, in the
Donnachie-Landshoff model \cite{DL} where the intercept of the Pomeron
trajectory  is $\alpha_{\pomeron}(0)=1.08$. However, the  hypothesis
of the Pomeron exchange implies the existence of the Pomeron
branch points in the plane of angular momentum which compete with
Pomeron exchange \cite{GribovPomeron}. Measurements of the total
inelastic diffraction cross section in $p\bar p$ scattering
\cite{Goulianos} discovered the  important role of shadowing effects which
can be described as due to the presence of
 the Pomeron branch points (for  recent
discussions of these issues see Refs. \cite{Kaidalov.etc}).
Account of the Pomeron branch points in the model with the value
of $\alpha_{\pomeron}(0)-1 > 0$ predicted blackening of
interaction at the range of  impact parameters $\bf{b} \propto \ln
(s/s_0)$ (see \cite{TerMartirosian} and references therein). This
corresponds to the Froissart-type behavior of the total cross
section in the $s\to \infty$ limit \cite{Froissart}.
Experimentally, at energies of the Tevatron partial amplitude for
$pp$ scattering at impact parameter ${\bf b}=0$ reaches a value
$\sim 0.95$.  Thus both theory and experimental data indicate that
the blackening of the soft interaction at particular fixed impact
parameters, $b$, until the unitarity limit is reached is one of
the distinctive features of soft hadron-hadron dynamics at high
energies.

On the other hand theoretical study  of DIS in perturbative QCD
(pQCD) \cite{Gross-Wilczek} and experimental investigations at
HERA \cite{HERA} found that the total cross sections of DIS, being
small, undergo a  significantly faster increase with energy than
the cross sections of soft hadron-hadron interactions. The
significant cross section of diffraction observed in DIS at HERA
 \cite{HERA} gives convincing evidence for the blackening of hard
QCD interactions at small $x$ \cite{FS}. In proton-proton
collisions at  collider energies, hard parton interactions  are
concentrated at significantly smaller impact parameters than
generic inelastic interactions \cite{FSW}. However, similar to the
DIS case, these hard parton interactions rapidly increase with
energy leading to blackening of interactions in the wide range of
central impact parameters. Thus it is important to analyze the
relative importance   of soft and hard QCD interactions in the
onset of the black body regime of high energy hadron collisions as
a function of the impact parameter.

In section 2 we analyze  the proton-proton  elastic scattering
amplitudes as a function of impact parameter ${\bf b}$. We argue
that the black regime originates from  hard interactions of
leading partons in the nucleons with the small $x$ gluon fields.
We use this to estimate the rate of  increase of the region in
impact parameter space where the interaction is black  and we find
that  it linearly increases  with $\ln s/s_0$. We also give an
alternative estimate based on the extrapolation of the soft
Pomeron dynamics which dominates at large ${\bf b}$ to the region
of smaller ${\bf b}$ and find that two estimates give similar
results at collider energies. We use these observations to argue
in section 3 that a universal pattern of blackening of the
interaction at fixed impact parameters leads to the universality
of cross sections in the limit of superhigh energies:
\begin{equation}
\sigma_{tot}(h_1,h_2)/\sigma_{tot}(pp)\to 1.\\
\label{uni1}
\end{equation}
Here $h_i$ can be a hadron (nucleus). The same result should be
valid for the structure functions of nuclei but at extremely small
$x$:
\begin{equation}
F_{2A}(x,Q^2)/F_{2N}(x,Q^2)\to 1.\\
\end{equation}
In the string models, similar universality arises as the
universality of the term  in the cross section which increases
linearly with energy - the "Pomeron" (loop contribution) related
to the universality of the graviton interaction \cite{Gurvich}.

It is worth mentioning here that universality of cross sections of
hadron-hadron, hadron-nucleus collisions at superhigh energies has
been discussed long ago. V.Gribov suggested universality of cross
sections of hadron-hadron, hadron-nucleus interactions
 \cite{Gribovuniversality} within the assumption that such cross
sections should become  constant at infinite energies. Later on,
the  universality of the Froissart limit for total cross sections
of hadron-hadron, hadron-nucleus and nucleus-nucleus collisions
was  suggested within a  particular generalized eikonal model of
the  supercritical Pomeron \cite{Karen}. This model gives
prescriptions for summing the unstable (divergent) series of terms
$\propto s^{n(\alpha_{\pomeron}-1)}$ due to multiPomeron exchanges
which rapidly increase with energy because of the intercept of the
bare supercritical Pomeron $\alpha_{\pomeron}(0)> 1$. There exist
also a number of eikonal models which combine elements of soft and
hard dynamics for all impact parameters, see \cite{engel} and
reference therein. However,  universality and the increase of the
radius of the hard interactions were not discussed in these
models. Note also, that the eikonal models neglect contribution of
the  enhanced Pomeron diagrams which  is rapidly increases with
energy and at superhigh energies becomes comparable with eikonal
diagrams \cite{TerMartirosyan}.

Our approach  assumes dominance of the Donnachie-Landshoff   soft
Pomeron exchange in peripheral collisions only. For the scattering
at central impact parameters, where according to preQCD Reggeon
Calculus strong interaction between the Pomerons is expected, our
approach accounts for the large cross section of hard processes
due to formation of large gluon densities with $p_{\perp}\approx 2
GeV$  and the consequent disappearance of soft elastic and
diffractive processes at energies of LHC and
above\footnote{Disappearance of diffractive processes for the
scattering at central impact parameters  has been observed
recently at FNAL. Remember also that in the black limit elastic
scattering occur at peripheral impact parameters only. In the
realistic situation inelastic diffraction arises from the
scattering at peripheral impact parameters where the interaction
is grey.}. Such nontrivial interplay of hard and soft dynamics
is
absent in preQCD multiPomeron exchange models. Note here, that 
 importance of hardinteractions in the
the Froissart limit in QCD follows from requirement of
the self consistency
 \cite{Kancheli}.

 We also analyze  how the increase of the  radius and strength of
the nucleon-nucleon interaction influences cross sections of
hadron-nucleus and nucleus-nucleus collisions at achievable
energies. To visualize the role of these effects we use the
formulae of the Glauber-Gribov model \cite{Glauber}\footnote{
V.Gribov has demonstrated that though the set of diagrams which
contributes at high energies is different from that accounted in
the original Glauber approach, the answer differs
only due to the contribution of the
inelastic diffraction - the inelastic shadowing corrections.
Relative contribution of these corrections is rather modest and
decreases at very high energies due to suppression of the
inelastic diffraction.}. We calculate the total and absorption
cross sections for pA collisions. The noticeable effects due to
the blackening of the nucleon-nucleon interaction can already be
observed at the LHC energies
 and above. The onset of black body limit leads to gradual
weakening of the A-dependence of cross sections which ultimately
results in the A-independent cross sections
corresponding to the universality regime of Eq. \ref{uni1}.

We evaluated the effective energy dependent radius of a nucleus
and estimated total cross sections for heavy ion collisions using
the popular Bradt-Peters expression \cite{Bradt}.  We compare this
result to that obtained with more refined Glauber-like model of
nucleus-nucleus interaction \cite{SatoBertsch,Kaidalovaa}. Also we
calculate
the distribution over the number of inelastic collisions in
nucleon-nucleus interaction \cite{Bertocci-Treleani}
and find that accounting for the energy dependent radius of a
nucleon-nucleon interaction
 leads to a
significant change of distribution. Analysis  of the possible role
of this effect in the interpretation of the heavy ion collisions
data in terms of the wounded nucleons already at energies of RHIC
and above is beyond the scope of this paper.

\section{Partial wave amplitudes for hadron-hadron collisions
at ultrahigh energies}

In this section we will use the properties of impact parameter
representation of the elastic scattering amplitude,
\begin{equation}
\Gamma ({\bf b},s)=\frac {1} {2i\pi k}\int d^2{\bf q} \exp [-i{\bf
q}\cdot {\bf b}]f_{NN}({\bf q},s), \label{pwave0}
\end{equation}
to discuss basic features of the superhigh energy $NN$ interactions.
Here $s$ is the invariant energy of $NN$ scattering, and  $\bf{q}$ is the
two-dimensional transverse momentum.

\subsection{Small impact parameter behavior of the hadron collision amplitude}

There are several generic features of $\Gamma ({\bf b},s)$ at
small impact parameters ${\bf b}$ which  can be derived from
unitarity of $S$ matrix,
from the current understanding of the spatial structure of the
fast nucleon and dynamics of hard interactions. Indeed, let us
consider nucleon-nucleon scattering at small impact parameters at
large $s$. An analysis, performed in Ref.\cite{FSW}, demonstrates
that
in this case the average transverse momenta of
partons from
nucleon become large after partons pass through the low $x$ gluon
fields of another nucleon. For example, at  Tevatron energies
quarks with $x\geq 0.2$ get average transverse momenta $\geq 1
GeV/c$. If a leading quark gets a transverse momentum $p_{\perp}$,
the probability for the nucleon to remain intact is roughly given
by the square of the nucleon form factor $F_N^2(p_{\perp}^2)$.
Since $F_N^2(p_{\perp}^2)\leq 0.1 $  for $p_{\perp}\geq 1 GeV/c$,
the probability of survival averaged over $p_{\perp}$ should be at
most $1/2$, provided average $p_{\perp}\geq 1 GeV/c$. Since there
are six leading quarks (plus a number of leading gluons)  {\it the
survival probability for two nucleons on small impact parameter},
$|1-\Gamma({\bf b},s)|^2$,  should go as a high power of the
survival probability for the case of one parton removal, $\approx
(1/2)^{6}$. Consequently,  $|1-\Gamma({\bf b}\sim 0,s)|^2$ is
close to 0 already at Tevatron energies.

Hence we conclude that
\begin{equation}
\Gamma ({\bf b}\sim  0,\sqrt{s}\geq ~2 ~TeV)\approx 1.
\label{pwave1}
\end{equation}
This evaluation is in a good agreement with the Tevatron data on
elastic $pp$ collisions and with their extrapolations
to the LHC energies.


Since the probability of hard inelastic interactions at fixed
impact parameter  increases with energy at least as the gluon
density at small $x$, $xG_N(x,Q^2)\propto x^{-n_h}$, the increase
of this probability should be proportional to $s^{n_h}$ with
$n_h\geq 0.2$.   The HERA data suggest that taming of the
interaction of small dipoles starts only when the probability of
inelastic interaction becomes large enough ($\geq 1/2$)). However
for such probabilities of single parton interactions the
mulitparton dynamics  insures the overall interaction to be
practically black. Hence, the multiparton interactions will ensure
a rapid (power law) onset of the regime of black interactions.

The analysis of the HERA data suggests \cite{FSW,FS02} that the
transverse distribution of the hard  partons (at the resolution
scale $p_t$) in nucleons can be described as $\propto \exp(
-m_h(x) b) $. Since  the interaction amplitude of the hard high
energy interaction is  $\propto s^{n_h}$ we find that the range of
${\bf b}\leq {\bf b}_F$ where the interaction is completely
absorptive should depend on energy as
\begin{equation}
 b_F \approx {n_h\ln {s\over s_0}\over m_h(x)}, \label{pwave4}
\end{equation}
which  corresponds to the Froissart limiting behavior. Note that
Eq.\ref{pwave4} is obtained in the limit of sufficiently high
energies when $m_h(x)$ for $x$ resolved at the corresponding
energy ($x\sim 4p_t^2/s$ where $p_t$ is hard scale) is much
smaller than that for the fast partons.

Actually, a few uncertainties limit
our accurate knowledge of the approaching to
the black body limit due to the mechanism of hard interactions.
First at all, there are the uncertainties in the $x$ dependence of
the gluon densities at very small $x$ where one needs to take into
account both $\ln (x_0/x), \ln (Q^2/Q_0^2)$ effects (for the
recent discussion see \cite{Altarelli-Ciafaloni}). Also one has
to account for the increase of the transverse spreading of the
gluon distribution with decrease of $x$. The small $x$
evolution is likely to lead to increase of $n_h$ at very small $x$
and sufficient virtualities. The neglected smearing of the fast
partons distribution leads to a decrease of effective mass
parameter at preasymptotic energies and to a somewhat faster
increase of $b_F$ with  energy.
Hence, both effects are likely increase the rate of the change of
$b_F$ at extremely small $x$.

The value of $m_h$ and rate of its decreasing with energy for
$x\leq 10^{-4}$ can be estimated based on the extrapolation
of $m_h$ extracted from the HERA data  for
$J/\psi$
photoproduction
which cover
$x\geq 10^{-4}$ and correspond to  $m_h(x=10^{-4}) \sim 0.75 GeV$:
\begin{equation}
m_{h}(s)=0.75\bigl [1-0.027\ln\bigl (\frac {s} {s_T} \bigr)\bigr ].
\label{mh}
\end{equation}
The expected limiting value  of $m_h$ is $2m_{\pi}$. Hence, with
the reasonable value $n_h=0.25$ we find that the true asymptotic is
likely to be reached at fantastically high energies $s\approx
10^{22}\, GeV^2$. At these energies $m_h$ will be the same for any
colliding hadrons build of light quarks.
 Note that in pQCD $n_h$ is determined by gluon
distribution and, hence, will be also universal, the same for any
colliding particles. In the case of hadrons with hidden heavy
flavor, the  onset of universality regime requires significantly larger
energies.

\subsection{Large  impact parameter behavior of the hadron-hadron amplitude}

To determine the behavior of the amplitude at large ${\bf b}$ we
can use arguments based on soft physics. At large ${\bf b}$
only single Pomeron interactions are possible as all multiPomeron
interactions have much smaller radius ($\sqrt{2}$ times smaller
for the  double Pomeron exchange, etc). Hence we can use here
information from the analysis of the data on pp scattering at
collider energies. Since we are interested in the large b behavior of
the amplitude which is determined by the properties of the
scattering amplitude at small $t$, we can use the simplest
exponential parameterization of the $t$ dependence of elastic
scattering, leading to the proton-proton amplitude due to the Pomeron
exchange
:
\begin{equation}
\Gamma_{\pomeron} ({\bf b},s)=
{\sigma (s)\over 4\pi
B_{\pomeron}(s)} exp\bigl [-{\bf b}^2/2B_{\pomeron}(s)\bigr ].
\label{pwave}
\end{equation}
We take the  amplitude to be  imaginary at high energies (the
ratio of real part of amplitude to the imaginary one at high
energies we consider here is practically constant and small
$\kappa \approx 0.1\div 0.13$). The parameter,
\begin{equation}
B_{\pomeron}=B_{0, \pomeron}+2\alpha_{\pomeron}'\ln(s/s_0)
\label{Pslope}
\end{equation}
is the slope of the $t$ dependence of the Pomeron exchange
contribution into the amplitude of elastic hadron-hadron collision
 \cite{Gribovslope,Chew}.  The discovery of the diffraction cone
shrinkage with increase of the energy in the elastic $pp$
collisions confirmed experimentally the energy dependence of
$B_{\pomeron}$  with the values of parameters
$B_{0,\pomeron}\approx 8.5\, GeV^{-2}, s_0=1\,GeV^2 $ and
$\alpha_{\pomeron} '=0.25\, GeV^{-2}$. The  Pomeron exchange
hypothesis predicts $B_{\pomeron}/B_{0,{\pomeron}}\approx 1.5$ at
the RHIC energies, $B_{\pomeron}/B_{0,{\pomeron}}\approx 2$ in the
kinematics of the LHC and $B_{\pomeron}/B_{0,\pomeron}\approx 2.7$
when energy is close to the GZK limit. Thus separate analysis of
peripheral and central collisions becomes appropriate at energies
of LHC and above. While the Pomeron model predicts that at
extremely high energies the partial amplitude at large impact
parameters  is determined by the Pomeron exchange
\cite{Gribovregge}, the situation is much more complicated at
smaller ${\bf b}$ because of shadowing effects (Pomeron branch
points). This will lead to renormalization of the Pomeron
parameters and to blackening of the interaction if
$\alpha_{\pomeron}(o)\geq 1$ cf. \cite{TerMartirosian}. So, the
calculation of the partial wave amplitude at non-peripheral ${\bf
b}$ is technically rather cumbersome. Instead, we assume (as
discussed in the Introduction) that partial amplitudes for ${\bf
b}\leq {\bf b}_F$ are $Im f({\bf b}\leq{\bf b}_F,s)=1$ and at
${\bf b}\geq {\bf b}_F$ are given by the dominant  single Pomeron
exchange\footnote{MultiPomeron cuts are decreasing with increase
of $b$ much faster than Pomeron exchange \cite{Gribovregge}. For
example, if $\Gamma_{\pomeron}({\bf b})\propto \exp(-\alpha {\bf
b}^2)$, n-Pomeron cut is $\propto \exp (-\alpha n {\bf b}^2)$.}.
 Then, the minimal
estimate of ${\bf b}_F$ can be obtained by  requiring the continuity
in the  matching of these two regimes:
$$
{\Gamma}_{\pomeron} ({\bf b}_F,s)= \frac {\sigma_{tot}^{pp}(s)}
{4\pi B_{\pomeron}(s)} \exp \bigl [-{\bf b}_F^2/{2B_{\pomeron}
(s)}\bigr ]=1.
$$
 For  $\sigma_{tot}^{pp}(s)$ we can use the
Landshoff-Donnachie parameterization of the Pomeron contribution
$\sigma_{tot}^{\pomeron} (s,s_0)=c\bigl [ {\frac {s} {s_0}}\bigr
]^{\alpha_{\pomeron}(0)-1}$ which provides a  good description
of the data in the region between ISR and Tevatron energies with parameter
$\alpha_{\pomeron}(0)-1=0.0808$. Hence,
we get the following energy dependence
from the condition that at the
energies of Tevatron, $s_T$, the partial wave of the $NN$ amplitude at
impact parameter $b=0$ becomes black ($\Gamma (b=0,s_T)=0.9\div
0.95\approx 1$):
\begin{equation}
{\bf b}_F^2\approx (\alpha_{\pomeron}(0)-1)\ln(s/s_T)B_{\pomeron}.
\label{cutoff}
\end{equation}
 At $s\to \infty$ the parameter
${\bf b}_F$ is universal-the same for any colliding particles:
$${\bf b}_F^2\approx 2\alpha_{\pomeron}' (\alpha_{\pomeron}(0)-1) \ln^2(s/s_T).$$

Another potentially important soft  contribution to the  partial wave
amplitude may arise from the
hadron scattering off
meson tails. Using the dispersion
representation of the amplitude over momentum transfer, $t$, it is
easy to obtain \cite{Gribovlectures}:
$$\Gamma({b})=cs^{\alpha(\mu^2)-1}exp [-\mu\, b].$$
The natural expectation
given the  fast decrease of the amplitude with increase of $b$ is that
$\mu$ is the minimal mass permitted in the channel with vacuum quantum
numbers i.e $\mu=2m_{\pi}$. Thus at energies when partial waves
with fixed impact parameter b become equal one we obtain:
$${ b}_F \approx
{1/\mu}(\alpha(\mu^2)-1)\ln(s/s_0).$$ This value is smaller than
that provided by the Pomeron exchange model, $${
b}_F=\sqrt{{B_{\pomeron}} (\alpha(0)-1)\ln(s/s_T)},$$ at achievable
energies {\bf but exceeds it at asymptotic energies}. The leading term
in ${\bf b}_F$ is the same if the partial amplitude is less than
one. It is easy to check that the combination of two types of the
${\bf b}$ dependencies discussed in this subsection does not
change the conclusion concerning the  universal value
for the leading term in ${\bf
b}_F$ at superhigh energies.

\subsection{Matching of small $b$ and large $b$ behavior}

\begin{figure}
\begin{center}
\epsfig{file=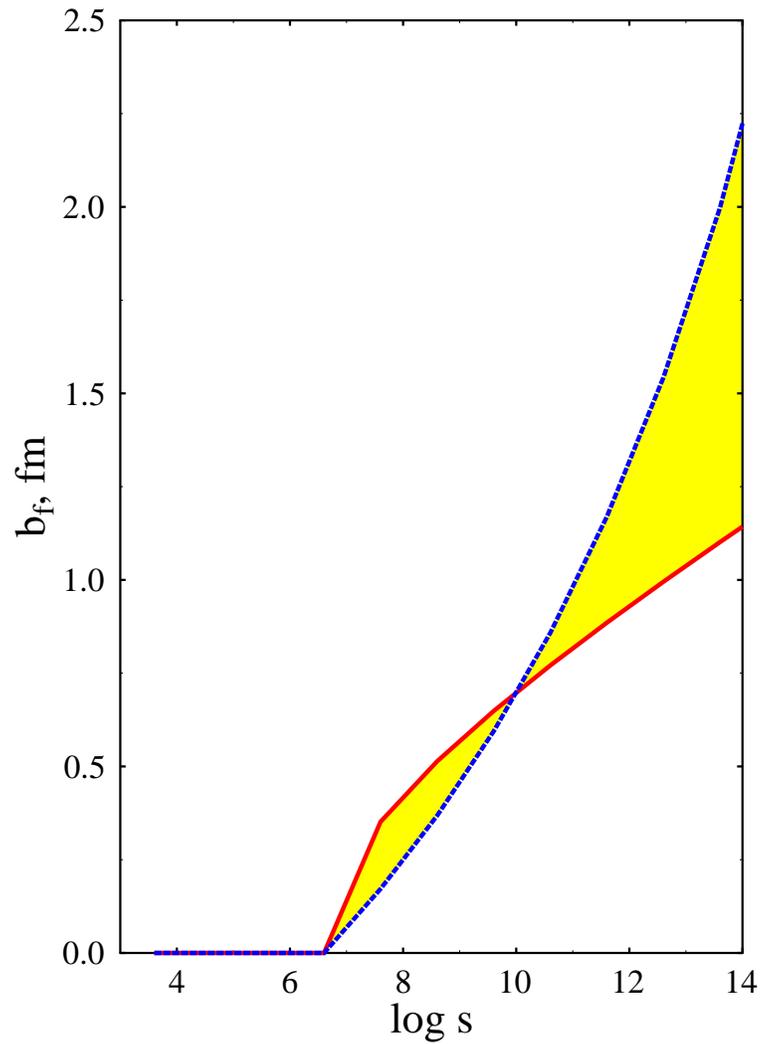, height=7in}
 \caption{Energy dependence of the black body regime cutoff parameter
   $b_F$. Solid red
 line - estimate based on the soft matching approximation. Dashed blue line -
 prediction of the hard matching approximation. Here and in the following
 figures ${\bf log \,s}$ stands for ${\bf log_{10}{(s/1\,GeV^2})}$.}
 \label{bfdep}
\end{center}
\end{figure}

We demonstrated above that the elastic amplitude is well
constrained both at small and large impact parameters. To build a
complete description we need to determine at what ${\bf b}_F$ the
two regimes match. Both soft and hard approximations give
practically the same result for ${\bf b}_F$ at energies in the
range from  the LHC  up to the GZK limit - see Fig.\ref{bfdep}.
This is due to the presence of a large constant term in
$B_{\pomeron}$. Hence we find that {\bf there is very good
consistency between the logic of matching starting from small $\bf
b$ and the logic of matching from large $\bf b$.} It means that
there is a smooth transition from hard regime at small ${\bf
b}\leq {\bf b}_F$ to soft regime at higher ${\bf b}\geq {\bf b}_F$
for the whole range of energies which maybe probed experimentally
at colliders and in cosmic ray interactions near  the GZK cutoff.
At the same time we find  that the asymptotic rate
of the increase of ${\bf b}_F$ is a factor of two larger in the
hard matching approximation. This indicates that at very high
energies, the  hard mechanism component of the black interactions
gives a dominant contribution to the cross sections. Also,
both hard and soft dynamics of the hadron-hadron interaction
predict the universal character of the ${\bf b}_F$ at the
superhigh energies.

\section{Universality of cross sections at extremely
large energies}

In the previous section we argued that  rate of the increase of
the size of
the region where $\Gamma({\bf b},s)\approx 1$ should be
followed by a rather steep drop of $\Gamma$ as given by  single
Pomeron exchange. Since the two scenarios of matching which we
considered in the previous section give practically identical
results for $10^3 \leq \sqrt{s} \leq 10^6 GeV$, we will consider here
 the soft matching dynamics which is obviously a more conservative way to estimate the approach to the
unitarity limit.

Imposing the condition of complete absorption at fixed ${\bf b}\leq
{\bf b}_F$ and using the Pomeron exchange formulae at larger ${\bf b}$,  we
can build the partial amplitude so that it includes the continuity
condition at the matching point
\begin{equation}
\Gamma_{NN} ({\bf b},s)=\Theta({\bf b}_F-{\bf b})+\exp\bigl
[-{({\bf b}-{\bf b}_F)}^2/2B_{\pomeron} \bigr ]\Theta({\bf b}-{\bf
b}_F). \label{profileNN}
\end{equation}

Since ${\bf b}_F$ is universal at superhigh energies the only
dependence on colliding particles is contained in the scale
$1/2B_{\pomeron}$ for the impact parameter distribution.

With the amplitude in  Eq.\ref{profileNN} one can calculate the total
cross section of the hadron-hadron interaction:
\begin{equation}
\sigma_{tot}=2\int \Gamma ({\bf b},s) d{\bf b}=2\pi (b_F^2
+2B_{\pomeron}).
\end{equation}
The slope of the $t$ dependence of the  elastic amplitude at $t=0$ is given
by the formula:
\begin{equation}
B={\frac {1} {\sigma_{tot}}}\int \Gamma ({\bf b},s) b^2 d{\bf b} =
{\frac {({b_F^4/2+b_F^2 2B_{\pomeron}+4B_{\pomeron}^2})} {2(b_F^2
+2B_{\pomeron})}}.
\end{equation}
So,
\begin{equation}
{\frac {\sigma} {4\pi
B}}=2(1-\frac{4B_{\pomeron}^2}{(b_F^2+2B_{\pomeron})^2+4B_{\pomeron}^2}).
\label{Bslope2}
\end{equation}
At  accelerator energies where $2B_{\pomeron}\gg b_F^2$   we
obtain relation: $\sigma/4\pi B \approx 1$ . In contrast, at
super high energies where  $b_F^2\gg 2B_{\pomeron}$ we obtain:
\begin{equation}
{\sigma/4\pi B}\approx 2(1-4B_{\pomeron}^2/b_F^4). \label{bound}
\end{equation}
Thus, at superhigh energies where $2\alpha' \ln(s/s_0)\gg
B_{0,\pomeron}$ the cross section and the slope of the t
dependence at $t=0$ become the same for all colliding particles.
The memory of the nature of colliding particles is lost as a
consequence of the blackening of the interaction and the ratio of
total cross sections for any colliding particles should be equal
to one:
\begin{equation}
\sigma_{tot}(h_1,h_2)/\sigma_{tot}(pp)\to 1.\\
\end{equation}

The same universality of structure functions is expected for
superhigh energies because the interaction and essential
impact parameters are increasing with energy and as the
consequence of the U matrix unitarity condition \cite{McDermott}:

\begin{equation}
F_{2 A}(x,Q^2)/F_{2 N}(x,Q^2)\to 1.
\end{equation}

A similar prediction  holds for the total cross section of
 a photon-nucleus scattering:
\begin{equation}
\sigma_{\gamma A}/\sigma_{\gamma p} \to 1.
\end{equation}

\section{Glauber model cross sections for pA and AA collisions }

\begin{figure}
\begin{center}
\epsfig{file=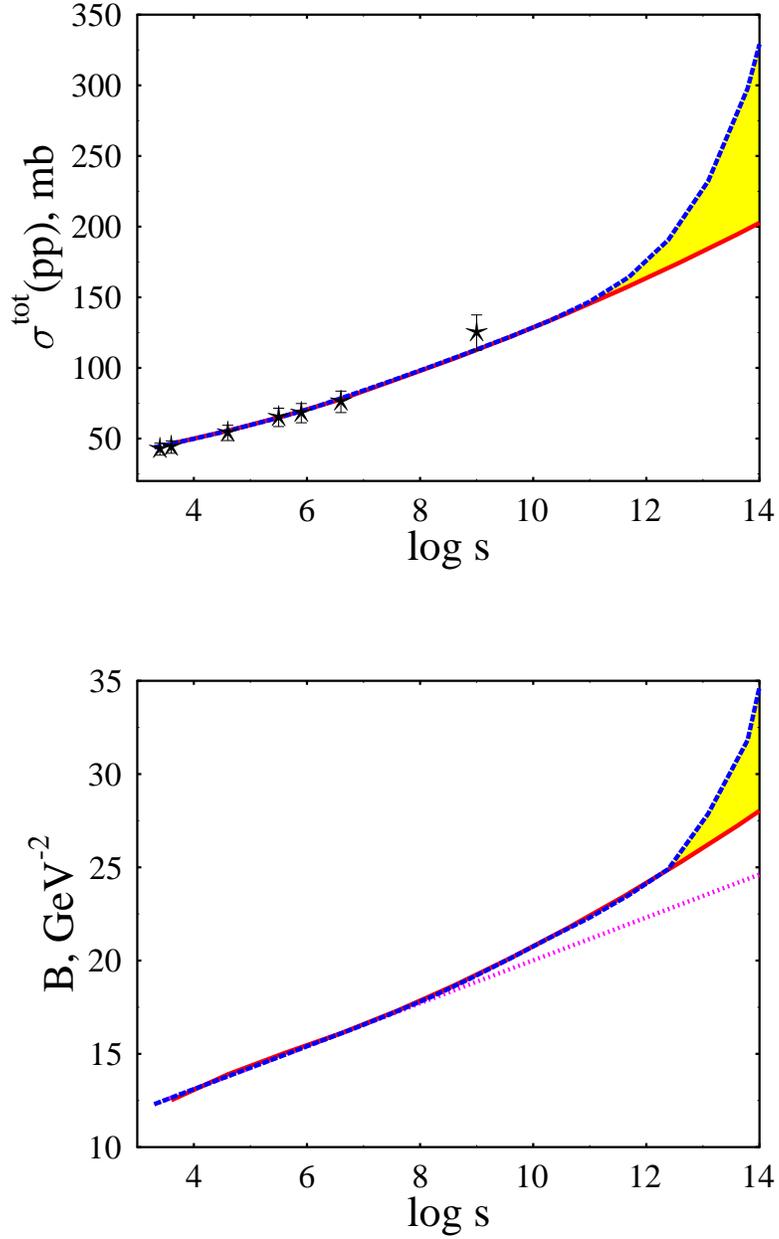, height=7in}
 \caption{Energy dependence of the total cross section and
 the slope parameter of $NN$ interaction. Solid line -
 the cross section and the slope parameter as dictated by the soft dynamics
 matching. Dashed line -hard mechanism of the blackening of $NN$ interaction.
 Dotted line presents the slope parameter $B_{\pomeron}$ due to the pomeron exchange}
\label{csb}
\end{center}
\end{figure}

To visualize physical phenomena related to increase of the radius
of a nucleon-nucleon interaction with energy and to learn whether
the trend toward universality can be observed at the energies
achievable at the accelerators or in cosmic ray we need to
parameterize the energy dependence of the elementary pp cross
section and the slope of t dependence for a  wide range of
energies. Up to the Tevatron energies ($s_T\approx 4\cdot 10^6
GeV^2$) we have chosen the parameterization of the total $pp$
cross sections in the form satisfying the Froissart theorem :
 \begin{equation}
\sigma_{tot}(pp)=\sigma_0(1+\epsilon\ln(s/s_0)+ \epsilon^2/2
\ln(s/s_o)^2) \label{sigma1}
\end{equation}
with $\epsilon=0.0808$ .  In the energy range where data are available this  form is almost identical to the one
suggested by  Donnachie and Landshoff.  The slope parameter
$B_{pp}(t=0)$ in this energy region is given by Pomeron exchange,
\begin{equation}
B_{pp}(t=0)=B_{\pomeron}=B_{0,\pomeron}+0.5\ln(s/1\,GeV^2).
\end{equation}
For  higher energies we use as input the profile function $\Gamma
({\vec b},s)$ which we built in the previous section
in order to calculate  the total, elastic
and inelastic cross sections and the slope parameter of the $NN$
amplitude. The matching of the cross section and $B_{NN}$ is
determined by using  $s=s_{T}$ as the reference point. The energy
dependence of the elementary NN cross section and of the slope
parameter $B_{NN}(s)$ is shown in Fig.\ref{csb}.

\begin{figure}
\begin{center}
\epsfig{file=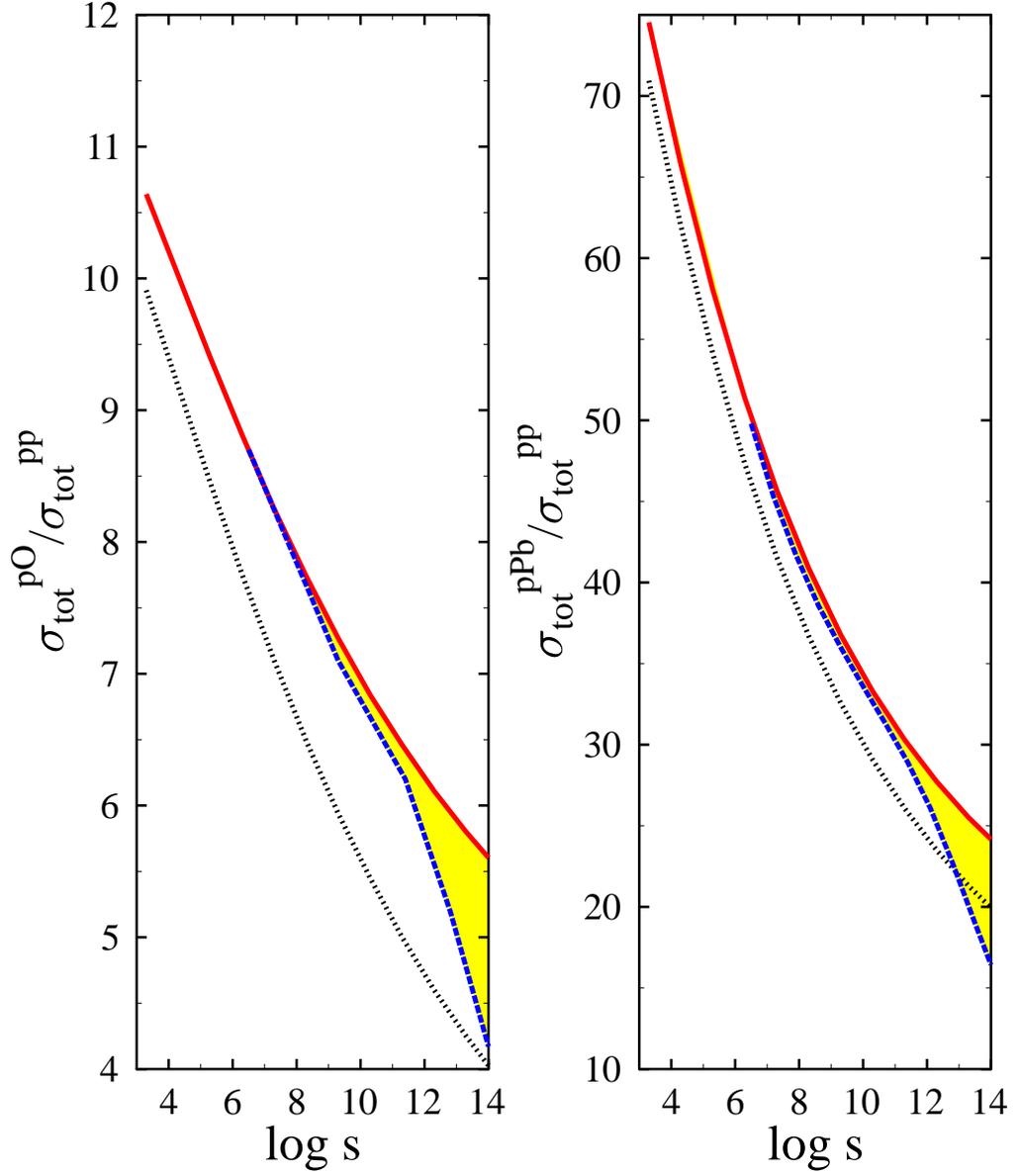, height=8in}  \caption{The dependence
of  $\sigma_{tot}(pA)/\sigma_{tot}(pp)$ on energy. Solid line -
the ratio calculated with $NN$ amplitude dominated by soft
dynamics, dashed line - hard mechanism of the blackening of
interaction. Dotted line is the ratio calculated neglecting by the
radius of interaction. } \label{energytcsrat}
\end{center}
\end{figure}
\begin{figure}
\begin{center}
\epsfig{file=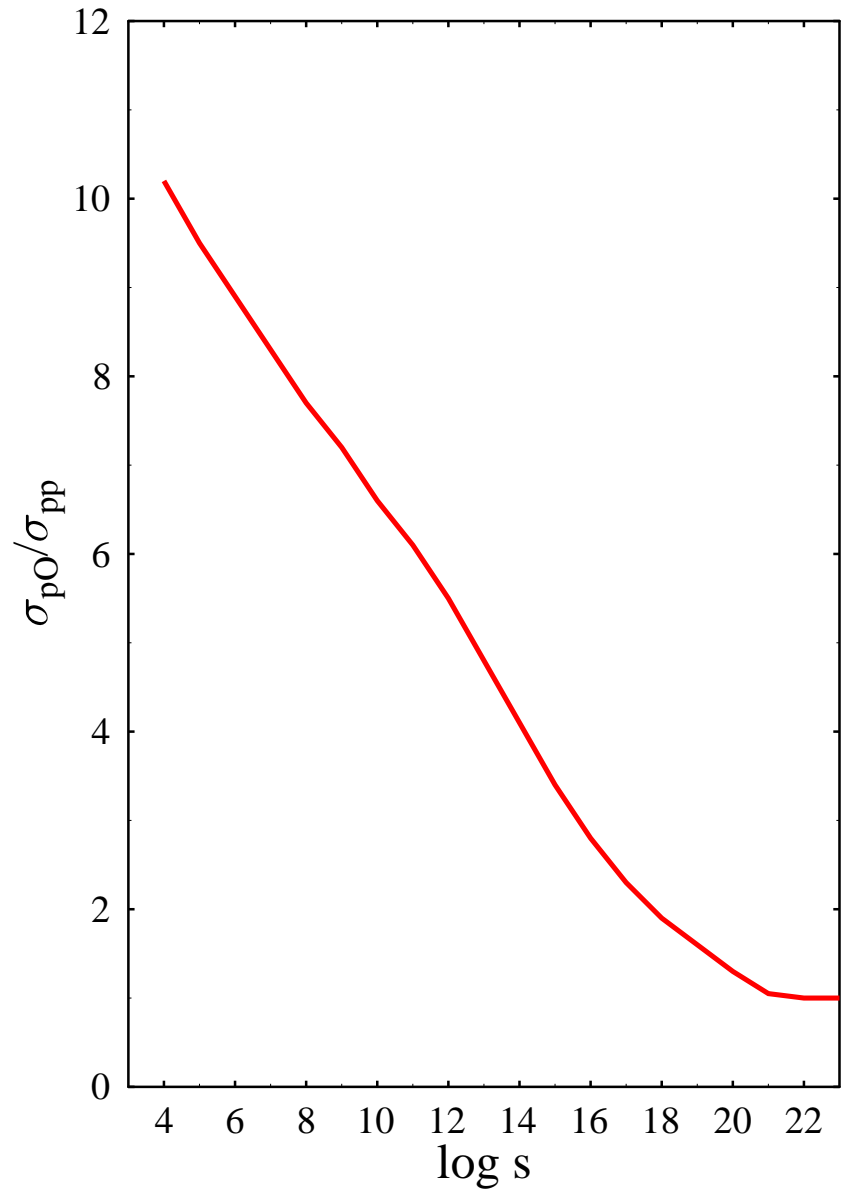, height=8in}
 \caption{The ratio of the proton-oxygen cross section to the proton-proton
 one
  as a function of energy. Calculation in the Gribov-Glauber model
  demonstrating the onset of the universality regime.}
\end{center}
\label{limit}
\end{figure}

\begin{figure}
\begin{center}
\epsfig{file=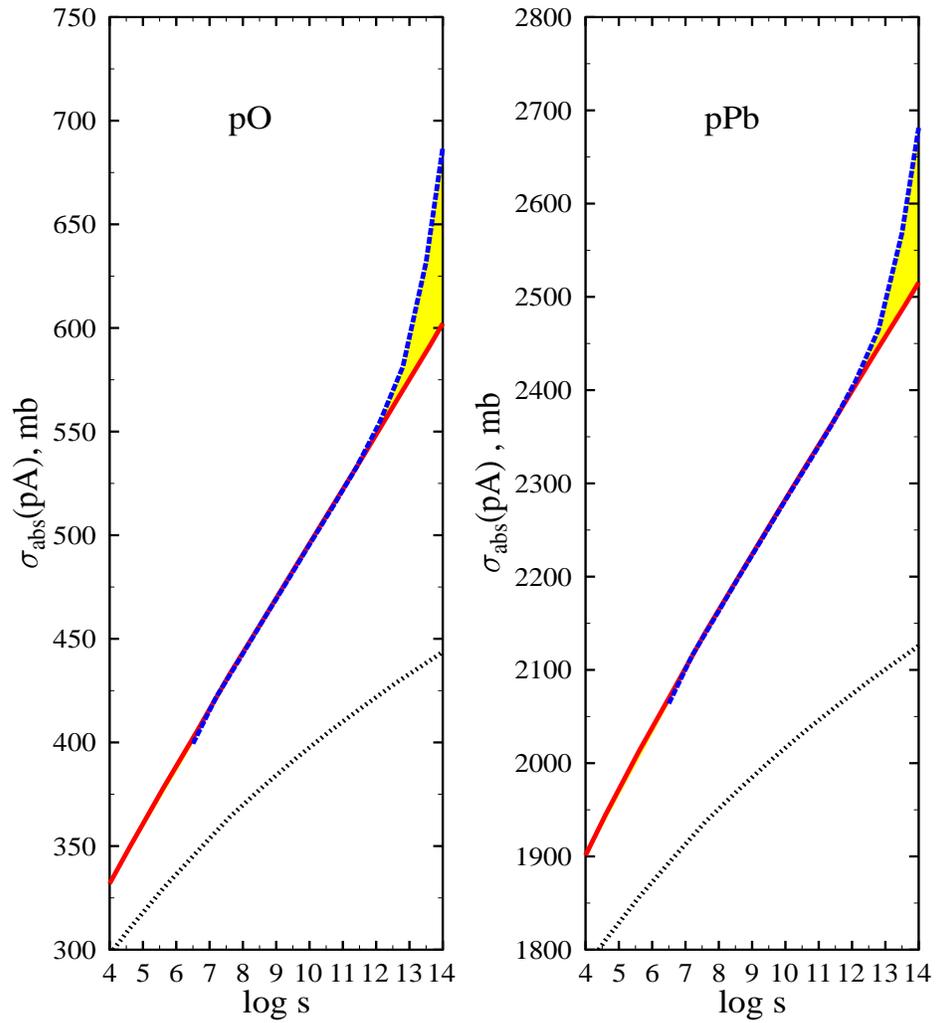, height=7in, width=5in} \caption{ The
absorption cross section of pA interaction as a function of
energy. Dashed line - calculation of cross section neglecting the
radius of interaction.} \label{absorbcs}
\end{center}
\end{figure}

 Now we can evaluate the  total
and absorption cross section for the proton-nucleus interaction at
high energies in the conventional Gribov-Glauber model:
\begin{equation}
\sigma_{tot}^{pA}(s)=2 \Re \int \limits_{0}^{\infty}\biggl
[1-\bigl [1-\int \rho (z,{\bf r}_{t})\Gamma_{NN} ({\bf b}-{\bf
r}_{t},s)d\,z\,d{\bf r}_{t}\bigr ]^A\biggr ]d{\bf b},\\
\end{equation}
and,
\begin{equation}
\sigma_{abs}^{pA}(s)=\int \limits_{0}^{\infty} \biggl [1-{ \left
\vert \biggl [1-\frac {2\sigma_{in}^{pp}} {\sigma_{tot}^{pp}}\int
\rho (z,{\bf r}_{t})\Gamma_{NN} ({\bf b}-{\bf r}_{t},s)d\,z\,d{\bf
r}_{t} \biggr ]^{A} \right \vert }^2 \biggr ] d{\bf b}.
\end{equation}
Here $\rho({\bf r})=A^{-1}\rho_{A}({\bf r})$ is the single nucleon
nuclear density normalized by the  condition $\int \rho({\bf
r})d{\bf r}=1$. We calculated the nuclear density $\rho_{A}$
within the Hartree-Fock model with the effective Skyrme
nucleon-nucleon interaction. Note that at intermediate energies
this nuclear model provides  a reasonable description of elastic
proton and electron scattering off nuclei along the periodical
table as well as the quasifree knockout of a  nucleon in (e,e'p)
and (p,2p) reactions without free parameters \cite{fsz}. In the
above formulae we neglected correlations between nucleons, single
and double inelastic diffraction. This is legitimate because for
central collisions, where these approximations look suspicious,
the  cross section is close to black limit and therefore
independent on details of the model. For the peripheral collisions
these effects are small.

The result of our calculations is shown in Fig.\ref{energytcsrat} as a
ratio of the total proton-nucleus cross section to the total cross
section of the $pp$ interaction as a function of the invariant energy
$s$ for  $p ^{16}O$ and $p ^{208}Pb$ collisions.
We present the ratio calculated  in the
approach of soft dynamics blackening(solid red line) and in the
hard regime (dashed blue line). The range between the two curves
can be treated as a measure of uncertainty of our approximation.  As
we already discussed,  both hard and soft regimes of blackening give
close results up to the energies of the GZK limit. At
higher energies,  the hard mechanism leads to 
a faster approach to
the  universality regime.
 For  comparison we also show
the ratio found in the model neglecting the
 radius of the $NN$ interaction(dotted line).
In the energy range where blackening is still a correction,
neglecting the radius of the interaction leads to a stronger
decrease of the $\sigma_{tot}^{pA}/\sigma_{tot}^{pp}$ ratio
in the case of light nuclei. This indicates a more important role
of peripheral interactions in the case of light nuclei.
Even so, the universal asymptotic is reached for light nuclei
at extremely high energies, see Fig.\ref{limit}.
The calculated absorption cross sections are shown in Fig.\ref{absorbcs}.

Accounting for the energy dependence of the radius of
hadron-hadron interaction in the energy domain where the $NN$
cross section becomes large and the radius of  the interaction becomes
comparable to the radius of the nucleus reveals  new effects
beyond those usually associated with Gribov-Glauber shadowing.
Scattering from the nucleus edge
and from the meson "halo" of a nucleus should lead to a decrease in the
dependence of the cross section on atomic number as compared to
nuclear shadowing effects. It is evident that at asymptotic
energies, where  the interaction is already black in the range of the
impact parameters considerably exceeding the radius of nucleus,  the
dependence on the atomic number in the nucleon-nucleus collisions
should completely disappear.

\begin{figure}
\begin{center}
\epsfig{file=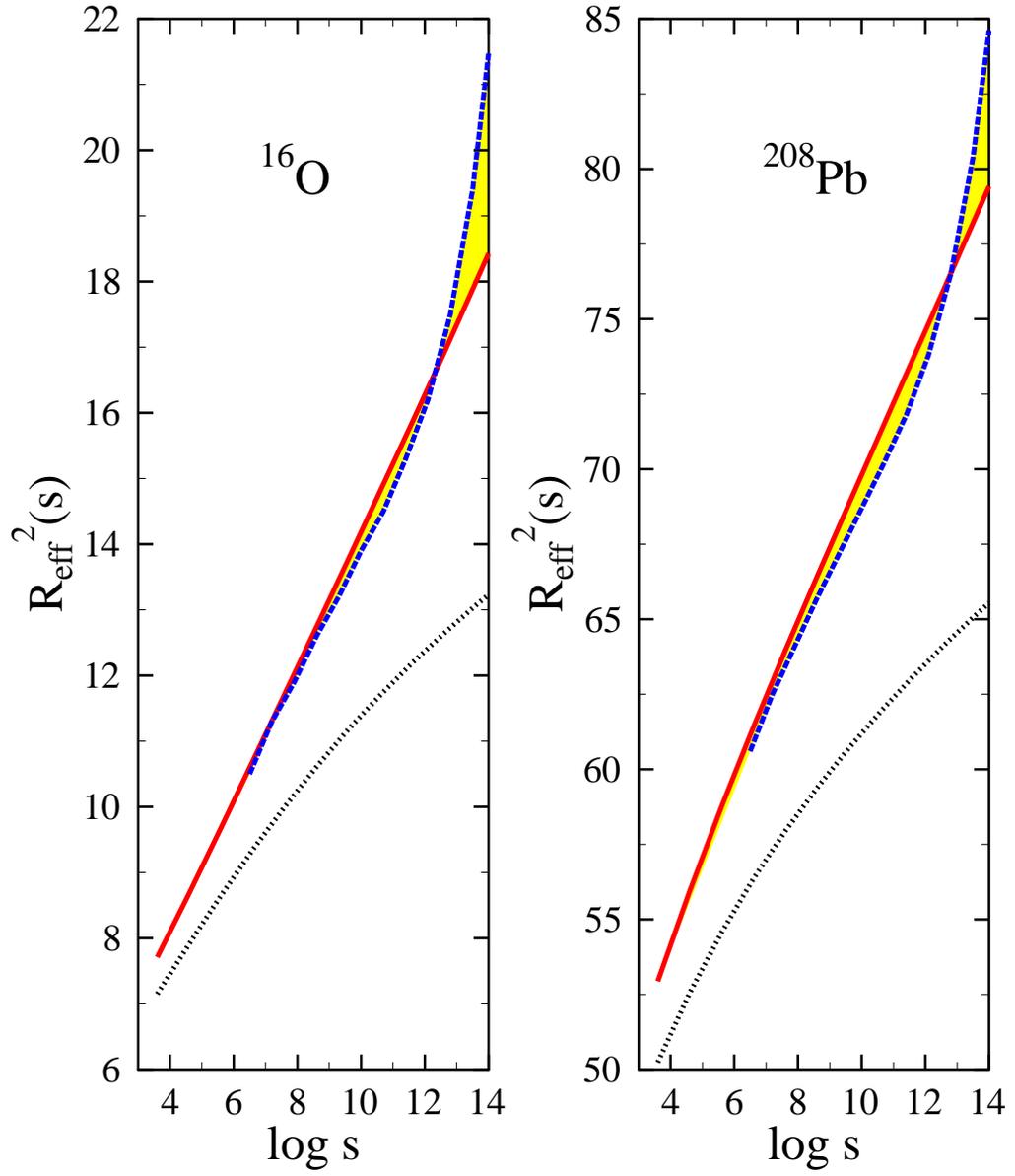, height=8in}  \caption{Increase with
energy of the effective nuclear radius in $pA$ collisions. Dotted
line - neglecting the radius of interaction. }
\label{effrad}

\end{center}
\end{figure}

\begin{figure}
\begin{center}
\epsfig{file=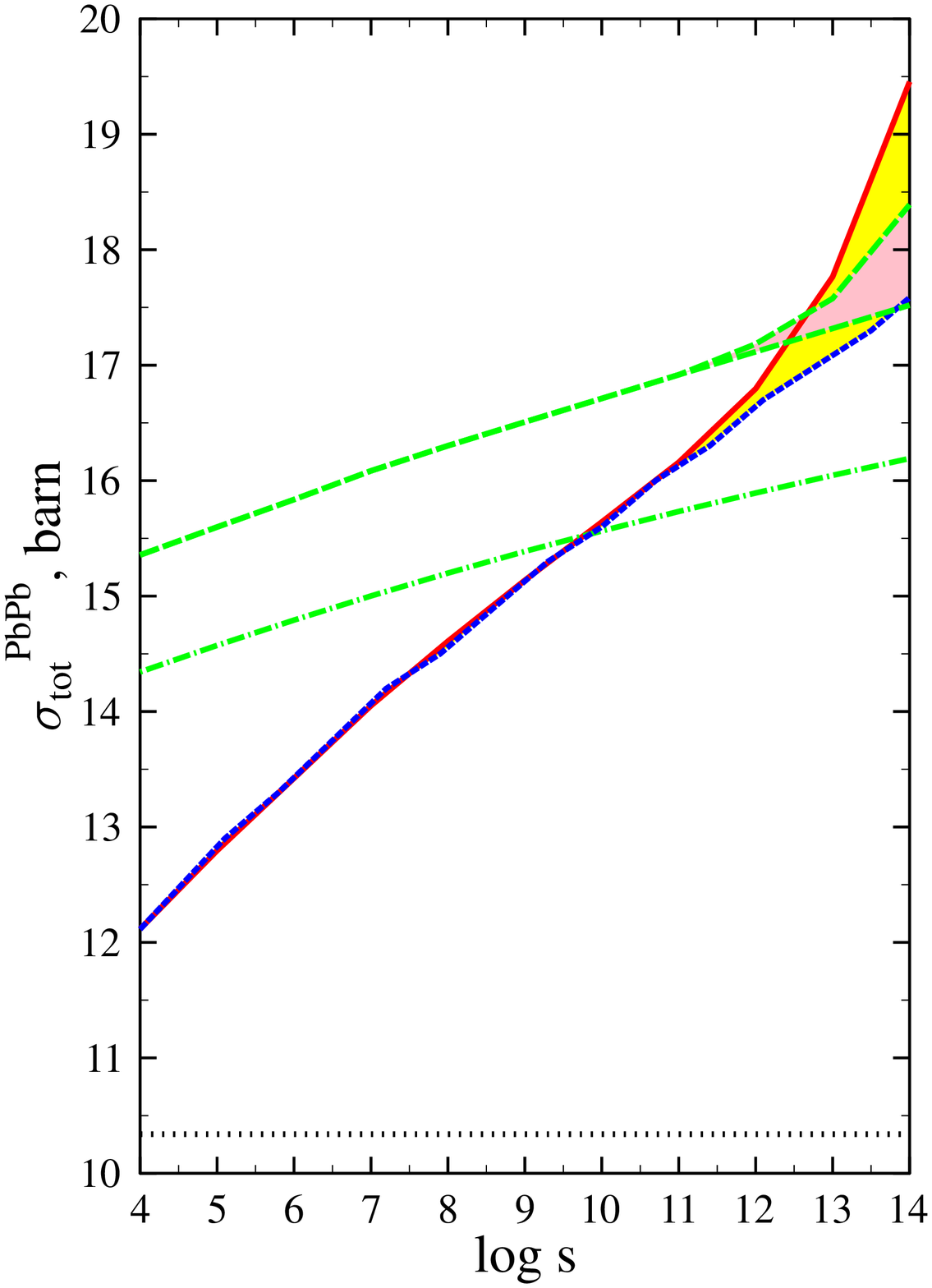, height=8in, width=5in} \caption{Cross
section of nucleus-nucleus collisions. Dotted line - the
Bradt-Peters cross section with static radii of nuclei; solid and
short-dashed lines - the Bradt-Peters cross sections with energy
dependent nuclear radii as provided by hard and soft dynamics
mechanisms correspondingly; dash-dotted line - optical Glauber
approach without accounting for the energy dependence of the
radius of $NN$ interaction; long dashed line - optical Glauber
model with accounting for the energy dependence of the interaction
radius } \label{bradt}
\end{center}
\end{figure}

We also estimated how the energy dependence of the $NN$
interaction radius in the high energy domain will affect the
total and inelastic nucleus-nucleus cross sections. These
quantities are used in the study of the relativistic heavy ion
collisions aimed to discover the new extreme state of the nuclear
matter - the Quark-Gluon plasma. The calculation of the
nucleus-nucleus cross sections in the Glauber-Gribov model is a
rather complicated problem. Instead, for the rough estimates of
the effect we used the generalization of the formulae of Bradt and Peters
\cite{Bradt} for the total cross section of scattering of two
heavy  ions:
\begin{equation}
\sigma_{tot}^{AB}=2\pi (R_{A}^{eff}(s)+R_{B}^{eff}(s)-c)^2.
\end{equation}
We use here the parameter
$c=0.8Fm$
as found by Bradt and Peters, and  we
calculate the cross section of heavy ion collisions using the
energy dependent nuclear radius $R_{eff}(s)$ determined from the
calculated proton-nucleus total cross section, $$R_{A}^{eff}(s)=
{\sqrt {\sigma_{tot}^{pA}(s)/2\pi}}.$$ The energy dependence of the
effective nuclear radius for lead is shown in
Fig.\ref{effrad}.
The energy dependence of  the $PbPb$ total cross
section is shown in Fig.\ref{bradt} where we also present the
cross section given by the Bradt and Peters formula with the static
nuclear radius(dotted line). The static nuclear radius was found
from the calculated HF nuclear density (if one uses the empirical
formula for the nuclear radius $R_{A}=r_{0}A^{2/3}$,  the HF density
of lead requires the value of $r_0=1.156$).  For comparison,  we
also calculated the total nucleus-nucleus cross section using the
simplified Glauber approach
expression \cite{SatoBertsch,Kaidalovaa}
\begin{equation}
\sigma_{AB}^{tot}=2\Re \int d {\bf b}\biggl [1-\exp \biggl (-
T_{AB}({\bf b})\biggr )\biggr ],
\end{equation}
where
\begin{equation}
T_{AB}({\bf b})=\int d {\bf b}_1 \int d {\bf b}_2 \Gamma_{NN}({\bf
b}_1 - {\bf b}_2 )\int dz_1 \rho_{A}({\bf b}_1 ,z_1 )\int dz_2
\rho_{B}({\bf b}-{\bf b}_2 ,z_2 ).
\end{equation}

We find significant corrections to the cross section of heavy ion
collisions calculated using the Bradt-Peters model with the static
nuclear radii already at  RHIC energies. We also compare in
Fig.\ref{bradt} these estimates of the total cross section to the
results of calculation within the simplified Glauber optical model
approach. In the range of energies $10^{4} \leq s \leq 10^{8}$ the
Bradt-Peters formula underestimates the cross section due to the
assumption of the sharp edges of nuclei, hence neglecting  the
contribution of the interaction of surface nucleons. However, at
 energies, close to the GZK limit the Bradt-Peters cross
sections with an account of  the energy dependence of the interaction
radius should be reasonable for the total cross section as well
as for the absorption cross section which is important for the
interpretation of cosmic ray data.
\begin{figure}
\begin{center}
\epsfig{file=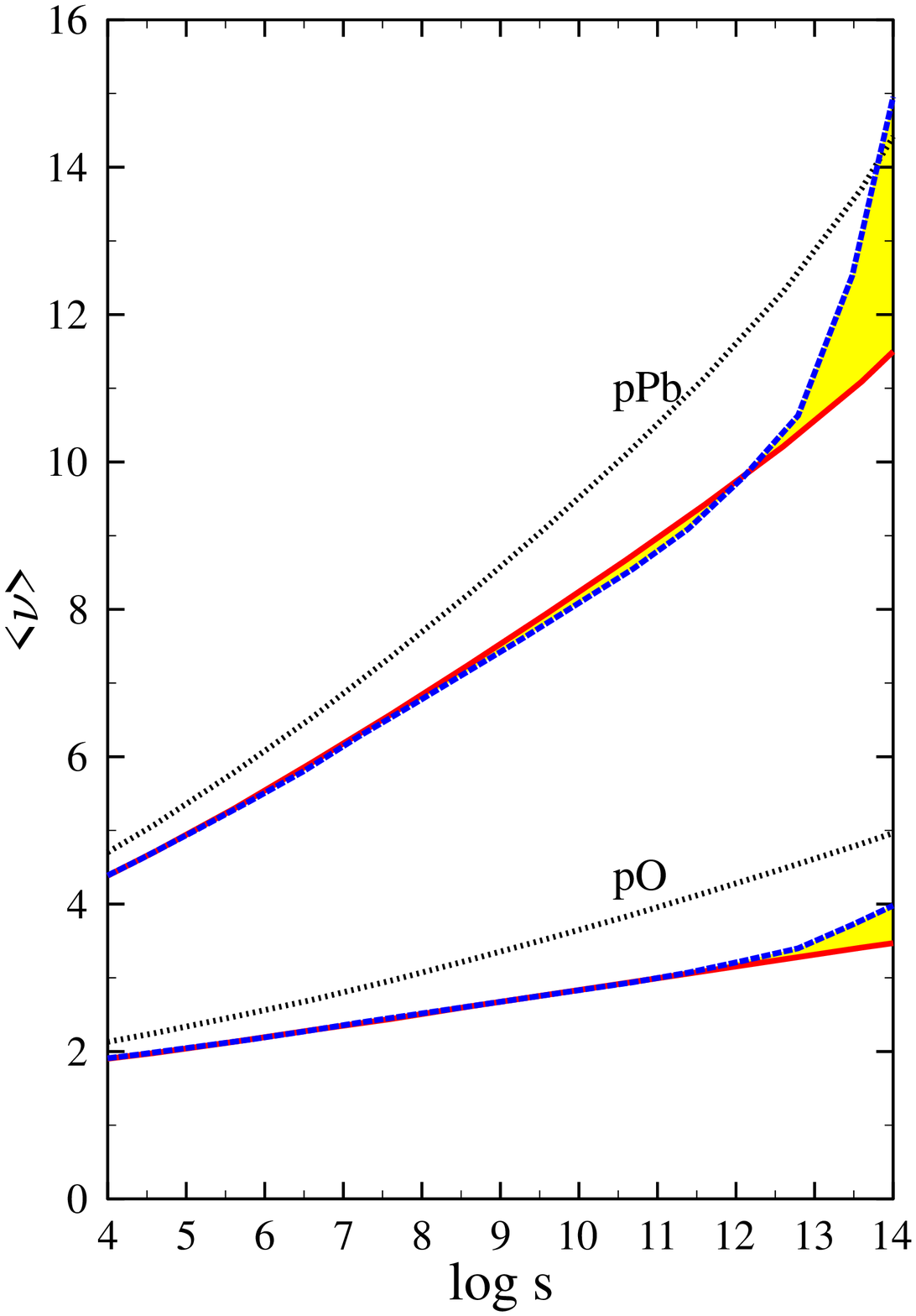, height=8in}
 \caption{ The average number of wounded nucleons in pA collision
  as a function of energy. Dotted line - calculation neglecting
  the energy dependence of the interaction radius.}
\label{wound}
\end{center}
\end{figure}

\begin{figure}
\begin{center}
 \epsfig{file=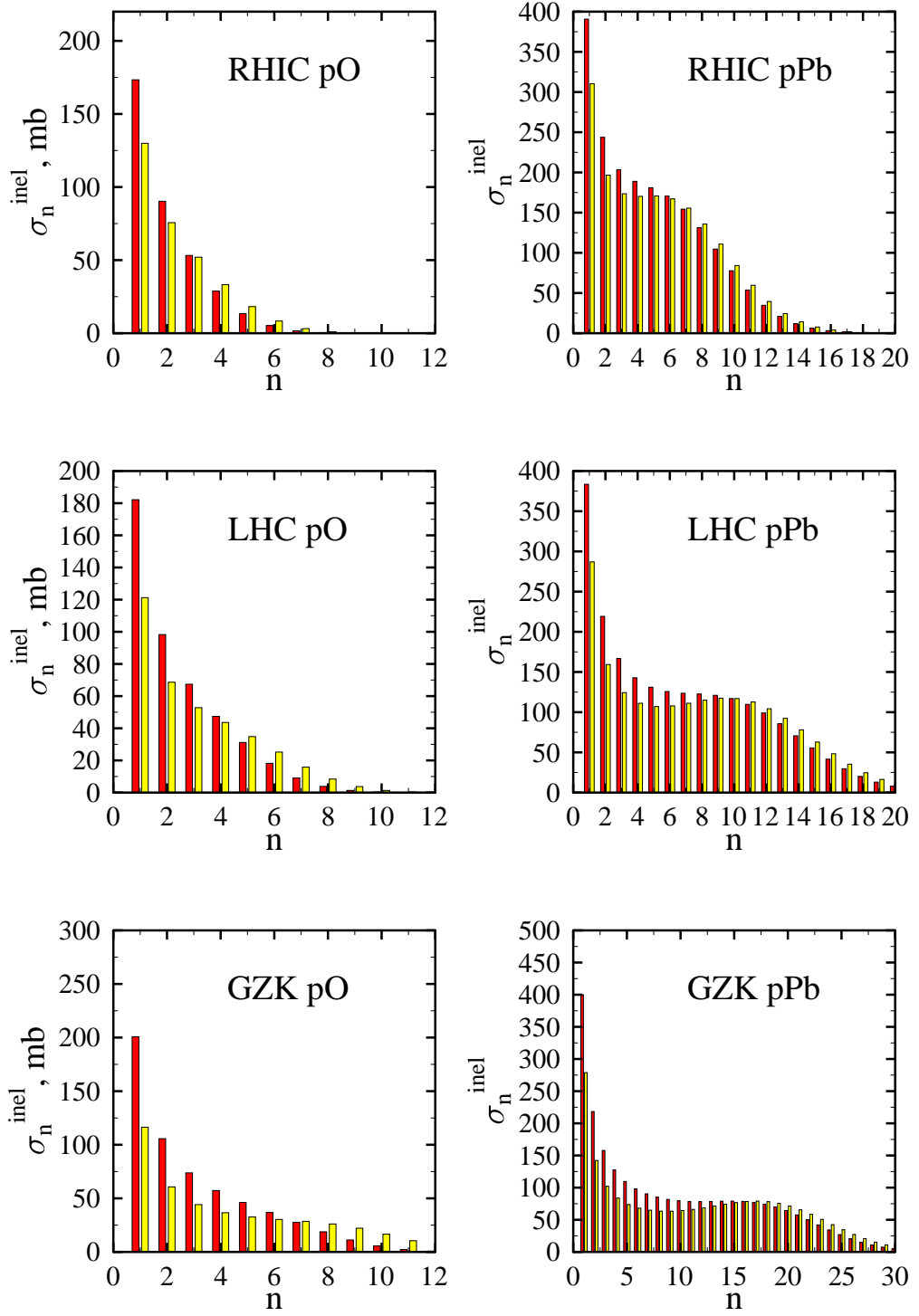, height=8in}
 \caption{The partial inelastic cross section as a function of the
 number of wounded nucleons. The yellow(light) bar is the cross section
 calculated neglecting the interaction radius.}
 \label{ndist}
\end{center}
\end{figure}

Now we want to demonstrate how accounting for the energy dependent
radius of the $NN$ interaction will change the distribution over
inelastic collisions with increasing energy. We calculated the
average number of wounded nucleons in nucleon-nucleus collisions
\cite{Bertocci-Treleani}
\begin{equation}
{\bar \nu}=\frac {A\cdot \sigma_{NN}^{in}(s)}
{{\sigma}_{pA}^{in}(s)},
\end{equation}
as a function of energy. Here,  the inelastic cross section is
calculated using the expression \cite{Bertocci-Treleani},
\begin{equation}
{\sigma}_{pA}^{in}(s)= \sum \limits_{n=1}^{A} \sigma_{n}(s),
\end{equation}
where the partial cross sections are given by formula
\begin{equation}
\sigma_{n}(s)=\frac {A!} {(A-1)!n!} \int d{\bf b}\biggl
[\sigma^{in}_{NN}(s)T({\bf b},s)\biggr ]^{n}\biggl
[1-\sigma^{in}_{NN}(s)T({\bf b},s)\biggr ]^{A-n} \label{partialc},
\end{equation}
with the generalized nuclear width function
\begin{equation}
T({\bf b},s)=\frac {2} {\sigma^{tot}_{NN}(s)}\int d{\bf r}_{t}
\Gamma_{NN}({\bf r}_{t}-{\bf b},s) \int dz \rho ({\bf r}_{t},z).
\end{equation}
 The energy dependence of the average number of wounded
nucleons calculated with and without  taking into account
the increase of the radius of the $NN$ interaction with energy is shown
in Fig.\ref{wound}.
The effect is still small, on the level of $10\%$,
at collider energies. At the same time,  at asymptotic energies where the
interaction becomes  black in a wide range of  impact
parameters  $<\nu>|_{s\to \infty} = A$ as a result of the
universality of the collision amplitudes. We also find
that,  though $<\nu>$ is weakly modified for collider energies,
the partial cross sections are affected much more strongly, see Fig.\ref{ndist}.
 In our calculations
we neglected corrections due to inelastic diffraction, by the
-momentum conservation law which are violated within the eikonal
approximation. With the  increase in radius of the nucleon-nucleon
interaction the basic assumptions leading to Gribov-Glauber
approximation, like
the neglect of the simultaneous interactions with two nucleons which
are located at different impact parameters, are violated
since the radius of the NN interaction becomes comparable to
the internucleon distance.

The impact of these effects on the interpretation of the RHIC and future LHC  data
requires separate analysis which is beyond this paper.

\section{Discussion and Summary}

We demonstrated that the cross section of $NN$
collisions at impact parameters growing as $\ln(s/s_0)$
 reaches the black limit due to the hard
dynamics. As a result, we argue that all hadronic and nucleus
cross sections become equal at ultra high energies. We analyzed
the role played by this effect for the total and absorption cross
sections of  the proton-nucleus collisions in a  wide energy
range. The effects are of  the order $10\%\div 20\%$  and, hence,
should be taken into account in the future analyses of the
precision LHC data. They may also be relevant for interpretation
of the cosmic ray data near the GZK limit. Dominance of hard
dynamics at small impact parameters both in the elastic amplitude
and in the structure of final states in the inelastic interactions
at small impact parameters \cite{FSW}  together with onset of the
universality of the cross sections can be considered as new
signals for a presence of the phase transition in NN interactions
at central impact parameters.

\section*{Acknowledgments}
This work has been supported in part by the USDOE and GIF.

\vfill\eject

\end{document}